\documentclass[12pt]{article}
\usepackage{amsmath,amssymb}

\newtheorem{thm}{Theorem} 
\newtheorem{cor}{Corollary}

\begin{document}

\author{Hironobu Kihara \\
Faculty of Science and Technology, Seikei University,\\
3-3-1 Kichijoji-Kitamachi,
 Musashino, Tokyo 180-8633, Japan\\
Research and Education Center for Natural Sciences,\\
Keio University,
4-1-1 Hiyoshi, Yokohama,
Kanagawa 223-8521, Japan\\
Faculty of Science, Ehime University, \\
10-13 Dogo-himata, Matsuyama, Ehime 790-8577, Japan}

\title{
Absence of Non-Trivial Supersymmetries and Grassmann Numbers in Physical State Spaces
}

\maketitle

\begin{abstract}
This paper reviews the well-known fact that nilpotent Hermitian operators on physical state spaces are zero,  thereby indicating that the supersymmetries and ``Grassmann numbers" are also zero on these spaces. 
 Next, a positive definite inner product of a Grassmann algebra is demonstrated,  constructed using a Hodge dual operator which is similar to that of differential forms. 
From this example, it is shown that the Hermitian conjugates of the basis do not anticommute with the basis and, 
therefore, the property that ``Grassmann numbers" commute with ``bosonic quantities" and 
anticommute with ``fermionic quantities", must be revised. 
 Hence, the fundamental principles of supersymmetry must be called into question. 
\end{abstract}

\clearpage

\section{Introduction}
~~In physics, materials in the universe are classified into bosons and fermions, 
which are described by quantum fields on the space-time. 
In order to explain fermions with classical theories, physicists use the ideal quantities ``Grassmann numbers". 
 However, 
few textbooks explain the domain of ``Grassmann numbers". 
Nevertheless, these ``Grassmann numbers" are treated as well-established mathematical objects in modern theoretical physics. 

It is difficult to find a discussion on the domain of the ``Grassmann numbers". 
The book ``Supermanifolds" by de Witt \cite{deWitt:1992} begins with a description of
 (infinite-dimensional) Grassmann algebras. 
According to his definition, a scalar field in a supersymmetric theory consists 
of infinitely many real scalar fields in the same sense as ordinary field theories, 
which sounds somewhat strange. 
As regards Lie algebra, commutation relations between bases, $L$, 
include all the information on the Lie algebra. 
Hence, physicists believe that the anti-commutation relations between $\theta_i$ define the algebra. 
However, in the case of ``Grassmann numbers", $\theta_i$ are not described 
as forming a basis but are instead referred to as ``parameters" or ``variables". 
In contrast, for a Lie algebra, $su(2)$, typically $\sigma_1,\sigma_2,\sigma_3$ 
are not referred to as ``Lie parameters". 
They are not parameters but invariable bases of the algebra. 
Thus, components and bases are confused here.  

The commutation relations between the generators of a Grassmann algebra 
and their Hermitian conjugates show that the appearance of Clifford algebras is 
natural in the unitary extension of Grassmann algebras. 
In addition, supersymmetries are imagined as symmetries whose ``parameters" are ``Grassmann numbers". 
The Grassmann algebra generated by an $n$-dimensional complex vector space has $\sum_{k=0}^n {}_nC_k = 2^n$ dimensions. 
Therefore, elements of the Grassmann algebra can be explained using these $2^n$ components. 
In addition, any vector spaces over a complex number field, $\mathbb{C}$, 
are vector spaces over a real number field, $\mathbb{R}$. 
For instance, $\mathbb{C}^n \simeq \mathbb{R}^{2n}$ are vector spaces over $\mathbb{R}$. 
Clearly, this statement is trivial, because $\mathbb{R}$ is a subfield of $\mathbb{C}$. 
Hence, elements of a Grassmann algebra generated by an $n$-dimensional complex vector space 
can be explained by $2^{n+1}$ real components. 
Such a realization tells us that if ``Grassmann numbers" are elements of a Grassmann algebra, the combination, $\epsilon Q$, of the ``Grassmann numbers", 
$\epsilon$, and ``infinitesimal generators of supersymmetries", $Q$, 
satisfy ordinary commutation relations without anti-commutation relations, 
and can be explained as a linear combination with real coefficients. 

In addition, many physicists hesitate to treat the ``Grassmann numbers" in terms of 
Grassmann algebra directly, which may be because of the confusion of components with bases.
The author  believes that the solid construction of theories is the most important 
aspect of theoretical physics, and therefore a comprehensive understanding of 
the ``Grassmann numbers" is required in order to accurately make use of this theory. 

As will be described below, through examination of the properties of Grassmann algebras and ``Grassmann numbers", 
the author suggests that the undefined tool ``Grassmann numbers" should be reconsidered. 
In the next section, 
it is proven that supersymmetries and ``Grassmann numbers" vanish on physical state spaces.
Then, in the third section, an example of commutation relations between the generators of 
a Grassmann algebra and their Hermitian conjugates is shown. 
Reconsideration of the property of the anti-commutation relations of ``Grassmann numbers" is proposed.
In addition, doubt is cast on the basic construction of supersymmetry.

\section{Nilpotent Hermitian Operators on Physical State Spaces}

~~ Hilbert spaces are complete vector spaces over $\mathbb{C}$ endowed with a Hermitian inner product. 
`Completeness' relates to the Cauchy sequence convergence property, however, this is not a central topic in this paper.  
A positive definite Hermitian inner product, represented by $\langle \cdot | \cdot \rangle$, is a Hermitian inner product which satisfies the following conditions: 
(i)~$ \langle \psi | \psi  \rangle \geq 0$ for all $\psi \in {\cal H}$, and
(ii)~$ \langle \psi | \psi  \rangle = 0 \Rightarrow \psi =0$. 
Hilbert spaces with positive definite inner products are called physical state spaces. 

Let us prove the nonexistence of nontrivial, nilpotent Hermitian operators on physical state spaces. 
\begin{thm}
Let ${\cal H}$ be a vector space over $\mathbb{C}$ endowed with a positive definite inner product, $\langle \cdot |  \cdot \rangle$. 
Suppose that $Y$ is a nilpotent Hermitian operator acting on ${\cal H}$, 
{\it i.e.} there is a natural number $n$, such that $Y^{n-1} \neq 0$ and $Y^n=0$, 
and $Y^{\dag}=Y$.
 Then, $Y=0$. 
\end{thm}

$\because$~~
Suppose that $n \geq 2$. $Y^{n-1} \neq 0$ and $Y^n =0$ implies that 
\begin{align}
&\langle Y^{n-1} \psi |  Y^{n-1} \psi \rangle = \langle Y^{n-2} \psi | Y^{\dag} Y^{n-1} \psi \rangle
= \langle Y^{n-2} \psi | Y^{n} \psi \rangle 
=0\ , \\
&\Rightarrow Y^{n-1} \psi=0 ~~(\forall\, \psi \in {\cal H})~.
\end{align}
Here, the first equality is obtained by taking the Hermitian conjugate and the second equality is supported by the Hermitian property, $Y^{\dag}=Y$. 
The third equality is obtained from the expression $Y^n=0$, and the final statement, $Y^{n-1} \psi=0$, is derived {from} the positive definite inner product. 
Because $Y^{n-1}\psi=0$ for every state, $\psi$, $Y^{n-1}$ must be zero. 
This contradicts the hypothesis and, therefore, we obtain $n=1$. Equivalently, $Y=0$. $\square$

Next, let us consider the operator $X$, where 
\begin{align}
X &= \epsilon_{\alpha} Q^{\alpha} + ( \epsilon_{\alpha} Q^{\alpha})^{\dag}~,
\end{align}
and $\alpha=1,2$. 
This kind of operator appears in the exponents of supersymmetries \cite{Wess:1992cp}.  
Suppose that $\epsilon_1$ and $\epsilon_2$ anticommute with each other, i.e.,  
$\epsilon_1 \epsilon_2 = - \epsilon_2 \epsilon_1$, and $\epsilon_1 \epsilon_1=\epsilon_2 \epsilon_2=0$. 
Usually, so-called ``Grassmann numbers" have the property that $\epsilon_{\alpha}$ commutes with ``bosonic quantities" and anticommutes with ``fermionic quantities". 
If we assume that the Hermitian conjugation transforms ``bosonic quantities" to ``bosonic quantities" 
and ``fermionic quantities" to ``fermionic quantities" 
(otherwise Hermitian conjugation is nothing but supersymmetry), then $\epsilon_{\alpha}$ should commute with $( \epsilon_{\alpha} Q^{\alpha})^{\dag}$.
From the ansatz of the anti-commutation, we can derive $\epsilon_{\alpha} \epsilon_{\beta} \epsilon_{\gamma}=0$, which indicates that $X$ is nilpotent. 
\begin{thm}
$X= \epsilon_{\alpha} Q^{\alpha} + ( \epsilon_{\alpha} Q^{\alpha})^{\dag}$ is nilpotent,  
where $\epsilon_{\alpha} \epsilon_{\beta} = - \epsilon_{\beta}\epsilon_{\alpha} $, 
$\epsilon_{\alpha} Q^{\beta} = - Q^{\beta} \epsilon_{\alpha}$, and
$\epsilon_{\beta} ( \epsilon_{\alpha} Q^{\alpha})^{\dag} = ( \epsilon_{\alpha} Q^{\alpha})^{\dag} \epsilon_{\beta}$. 
\end{thm}
$\because$
\begin{align}
X^5 &= (\epsilon_{\alpha} Q^{\alpha} + ( \epsilon_{\alpha} Q^{\alpha})^{\dag})^5\ , \\
&= (\epsilon_{\alpha} Q^{\alpha})^5\ , \cr
&+  (\epsilon_{\alpha} Q^{\alpha})^4 ( \epsilon_{\beta} Q^{\beta})^{\dag}
+ (\epsilon_{\alpha} Q^{\alpha})^3 ( \epsilon_{\beta} Q^{\beta})^{\dag} (\epsilon_{\gamma} Q^{\gamma})
+ \cdots  + ( \epsilon_{\alpha} Q^{\alpha})^{\dag}(\epsilon_{\beta} Q^{\beta})^4\ , \cr
&+ \cdots\ , \cr
&+ \left\{ (\epsilon_{\alpha} Q^{\alpha})^5 \right\}^{\dag}~ \ .
\end{align}
Here, the term $(\epsilon_{\alpha_1} Q^{\alpha_1}) (\epsilon_{\beta_1} Q^{\beta_1})^{\dag}
(\epsilon_{\alpha_2} Q^{\alpha_2}) (\epsilon_{\beta_2} Q^{\beta_2})^{\dag}
(\epsilon_{\alpha_3} Q^{\alpha_3})$ vanishes because it contains three $\epsilon$'s. Then
\begin{align}
&(\epsilon_{\alpha_1} Q^{\alpha_1}) (\epsilon_{\beta_1} Q^{\beta_1})^{\dag}
(\epsilon_{\alpha_2} Q^{\alpha_2}) (\epsilon_{\beta_2} Q^{\beta_2})^{\dag}
(\epsilon_{\alpha_3} Q^{\alpha_3})\  \cr
&= - \epsilon_{\alpha_1} \epsilon_{\alpha_2}\epsilon_{\alpha_3} 
 Q^{\alpha_1} (\epsilon_{\beta_1} Q^{\beta_1})^{\dag} Q^{\alpha_2}
 (\epsilon_{\beta_2} Q^{\beta_2})^{\dag} Q^{\alpha_3}~\ .
\end{align}
In addition, each $X^5$ term includes factors $\epsilon_{\alpha} \epsilon_{\beta} \epsilon_{\gamma}$ 
or $(\epsilon_{\alpha} \epsilon_{\beta} \epsilon_{\gamma})^{\dag}$, and therefore $X^5=0$. $\square$\\

Theorem 2 is the direct conclusion of the property 
`$\epsilon_{\alpha}$ commutes with ``bosonic quantities" and 
anticommutes with ``fermionic quantities" '.
\begin{cor}
\rm (i)~ $X=0$. ~~(ii)~~
$\epsilon Q=0$. 
\end{cor}
$\because$~~
(i)~~ It is obvious from the definition that $X=X^{\dag}$. 
Therefore, $X$ is a nilpotent Hermitian operator and, from theorem 1, we conclude that $X=0$. 
(ii)~~$X=0$ is equivalent to $\epsilon Q = - (\epsilon Q)^{\dag}$. 
Again, $i \epsilon Q$ is a nilpotent Hermitian operator and $\epsilon Q=0$. $\square$

Corollary 1 raises a question not only about supersymmetries but also concerning ``Grassmann numbers". 
\begin{cor}
If `$\epsilon$ commutes with ``bosonic quantities" and anticommutes with ``fermionic quantities"', then 
$\epsilon=0$ on the physical state spaces. 
\end{cor} 

$\because$~~Assume that $\epsilon$ satisfies the condition; 
`$\epsilon$ commutes with ``bosonic quantities" and anticommutes with ``fermionic quantities"'. 
$Z= \epsilon + \epsilon^{\dag}$ is a nilpotent Hermitian operator and we therefore obtain  
$Z=0$, while $\epsilon^{\dag}+ \epsilon=0$ shows that $i \epsilon$ is nilpotent and Hermitian. 
Hence, $\epsilon=0$ is required. $\square$

Therefore, in order to treat ``non-trivial Grassmann numbers", 
ghost states, which are defined as states whose existence probabilities are negative, are required.
As it is apparent that ``Grassmann numbers" strongly 
depend on the  representation ${\cal H}$, the fundamental principles of supersymmetry are called into question.
In order to examine the difference between ``Grassmann numbers" 
and elements in the unitary extension of Grassmann algebras, a basic example of a 
Grassmann algebra generated by a two-dimensional 
vector space over $\mathbb{C}$ is given in the next section.

\section{Hermitian conjugation of a Grassmann algebra}

~~Let $V$ be a two-dimensional vector space over $\mathbb{C}$, $V \simeq \mathbb{C}^2$, 
and suppose that $V$ is endowed with a positive definite Hermitian inner product, $\langle \cdot | \cdot \rangle$. 
An orthonormal basis, $\{e_1,e_2\}$, is fixed in this section, such that  
$\langle e_i | e_j \rangle = \delta_{ij}$.  
Now, the Hermitian inner product can be expressed using the expansion coefficients with respect to the basis. 
Suppose that two vectors, $v,w \in V$, are expanded with respect to $\{e_1,e_2\}$, such that 
\begin{align}
 v &= v_1 e_1 + v_2 e_2 = \begin{pmatrix} e_1 & e_2 \end{pmatrix} 
 \begin{pmatrix}v_1\\v_2\end{pmatrix} ~,~~
 w = w_1 e_1 + w_2 e_2
 = \begin{pmatrix} e_1 & e_2 \end{pmatrix} 
 \begin{pmatrix}w_1\\w_2\end{pmatrix}~.
\end{align}
Then, the inner product of $v$ and $w$ 
can be written as 
\begin{align}
 \langle v | w \rangle &= \langle  v_1 e_1 + v_2 e_2 |  w_1 e_1 + w_2 e_2 \rangle\ , \\
 &= v_1^* w_1  \langle e_1 | e_1 \rangle 
 + v_1^* w_2  \langle e_1 | e_2 \rangle 
 + v_2^* w_1  \langle e_2 | e_1 \rangle 
 + v_2^* w_2  \langle e_2 | e_2 \rangle\ , \\
 &= v_1^* w_1 + v_2^* w_2 = \begin{pmatrix} v_1^* & v_2^* \end{pmatrix} 
 \begin{pmatrix} w_1 \\ w_2 \end{pmatrix}\ .
\end{align}
Next, we wish to examine the Grassmann algebra, $A$, generated by $V$. 
The algebra $A$ is a vector space over $\mathbb{C}$.  
The product in $A$ is denoted by the wedge product, $\wedge$, and, 
thus, the product of $v,w \in A$ is represented by $v \wedge w$. 
The multiplication is bilinear and satisfies the associative and distributive laws. 
In addition, the basis of $A$ can be constructed from the basis of $V$, so that  
 $\{1 , e_1 , e_2 , e_1 \wedge e_2 \}$ form a basis of $A$. Note that
$\omega = e_1 \wedge e_2$ is called the volume form. 
 
The multiplication relations between these bases can then be expressed as 
\begin{align}
1 \wedge e_i &= e_i \wedge 1 = e_i~,& 
1 \wedge ( e_1 \wedge e_2) &=  ( e_1 \wedge e_2 ) \wedge 1 = e_1 \wedge e_2~,\\
e_i \wedge e_j &= - e_j \wedge e_i~,&
e_i \wedge ( e_1 \wedge e_2) &= (e_1 \wedge e_2) \wedge e_i = 0~\ ,
\end{align}
and a positive definite inner product on $A$ can be defined. 

Let us consider the complex conjugation operator $C$ with respect to the basis, 
$\{ 1 , e_1  , e_2 , e_1 \wedge e_2 \}$. $C$ is a real linear operator and, 
for $X= v_0 + v_1 e_1 + v_2 e_2 + v_3 e_1 \wedge e_2 \in A$, 
\begin{align}
 C (X) &= v_0^* + v_1^* e_1 + v_2^* e_2 + v_3^* e_1 \wedge e_2 ~. 
\end{align}

The Hodge dual operator $*$ is a complex linear operator on $A$ and 
the action on the basis $\{1 , e_1 , e_2 , e_1 \wedge e_2 \}$ is defined as  
\begin{align}
* 1 & =e_1 \wedge e_2 ~,& 
* e_1 &= e_2 ~,& *e_2 &= - e_1 ~,&  
* e_1 \wedge e_2 &= 1~.
\end{align}
Now, let $X \in A$ and $X= v_0 +  v_1 e_1 + v_2 e_2 + v_3 e_1 \wedge e_2$. Then,
\begin{align}
* X &= v_0 e_1 \wedge e_2 + v_1 e_2 - v_2 e_1 + v_3\ .
\end{align}
If $X,Y \in A$ are expanded with respect to the previous basis, we have 
\begin{align}
X &= v_0 + v_1 e_1 + v_2 e_2 + v_3 e_1 \wedge e_2 ~,\\
Y &= w_0 + w_1 e_1 + w_2 e_2 + w_3 e_1 \wedge e_2  ~. 
\end{align}
It is apparent that the term proportional to the volume form in $C(X) \wedge * Y$ gives a Hermitian form on $A$, and hence 
\begin{align}
C(X) \wedge  * Y %&=  \left( v_0^*  + v_1^*  e_1 + v_2^*  e_2 + v_3^* e_1 \wedge e_2  \right) \wedge 
%\left( w_0 e_1 \wedge e_2 + w_1 e_2 - w_2 e_1 + w_3 \right) \\
&=  v_0^* w_3\ , \cr
&+ v_0^* w_1 e_2 - v_0^* w_2 e_1 + w_3  v_1^*  e_1 +  w_3 v_2^*  e_2\ , \cr
&+ (v_0^* w_0   + v_1^*  w_1  +  v_2^* w_2  + v_3^* w_3 ) e_1 \wedge e_2~,\\
\langle X | Y \rangle & \equiv v_0^* w_0 +  v_1^* w_1 +  v_2^* w_2  +  v_3^* w_3  ~.
\label{eqn:defofinner}
\end{align}
As an alternative explanation, 
$ \langle X | Y \rangle =  * [ \omega \wedge * \{ C(X) \wedge *Y \} ]$. 
The definition given above (Eq.\ref{eqn:defofinner}) is independent of the choice of orthonormal basis $\{e_1,e_2\}$, 
and every unitary transformation represented by $U$ yields a new orthonormal basis, $(f_1 ~ f_2) = (e_1 ~ e_2) U$. 
Here $U$ is a $2 \times 2$ unitary matrix and $U^{\dag} U= U U^{\dag}=1$. 
Of course, we now have $C(f_i) \neq f_i$  in general and
$1$, a base of $A$, is invariant under the transformation by $U$. 

The determinant of $U$ is a complex number with a unit norm, 
$|U| = e^{i \alpha}$, where $\alpha$ is a real number. 
Let us rewrite $X$ in terms of the new basis, with 
\begin{align}
X &= v_0 + v_1 e_1 + v_2 e_2 + v_3 e_1 \wedge e_2\ , \\
&= v_0' + v_1' f_1 + v_2' f_2 + v_3' f_1 \wedge f_2 ~.
\end{align}
Then, the relationships between the coefficients can be expressed as
\begin{align}
v_0' &= v_0~,& 
 \begin{pmatrix}v_1' \\v_2' \end{pmatrix} &= U^{-1} \begin{pmatrix}v_1 \\v_2 \end{pmatrix}~, & 
v_3'&= e^{-i \alpha} v_3~.
\label{eqn:relcoeff}
\end{align}
Here, Eq. \ref{eqn:relcoeff} implies that the previous Hermitian inner product is preserved under 
the transformation by $U$, and therefore 
\begin{align}
&(w_0')^* v_0' + (w_1')^* v_1' + (w_2')^* v_2' + (w_3')^* v_3' \cr
&= w_0^* v_0 + \begin{pmatrix} w_1^* & w_2^* \end{pmatrix} U U^{-1} \begin{pmatrix} v_1 \\ v_2 \end{pmatrix} 
+ w_3^* e^{i \alpha} e^{-i \alpha} v_3~\ ,\\
&= w_0^* v_0 + w_1^* v_1 + w_2^* v_2 + w_3^* v_3 ~~.
\end{align}

Next, let us consider the representation of $A$ on itself. 
Assume that the representation $R : A \rightarrow {\rm End}_{\mathbb{C}} (A)$ is defined by 
$R(X) Y := X \wedge Y$. 
For instance, 
\begin{align}
R(1) X &= 1 \wedge X=X\ , \\
&= \begin{pmatrix} 1 & e_1 & e_2 & e_1 \wedge e_2 \end{pmatrix}
\begin{pmatrix} 
1 & 0 & 0 & 0 \\
0 & 1 & 0 & 0 \\
0 & 0 & 1 & 0 \\
0 & 0 & 0 & 1 
\end{pmatrix}
\begin{pmatrix} 
v_0 \\ v_1 \\ v_2 \\ v_3 
\end{pmatrix}
~,\\
R(e_1) X &= e_1 \wedge X = v_0 e_1 + v_2 e_1 \wedge e_2\ , \\
&=\begin{pmatrix} 1 & e_1 & e_2 & e_1 \wedge e_2 \end{pmatrix}
\begin{pmatrix} 
0& 0 & 0 & 0 \\
1 & 0 & 0 & 0 \\
0 & 0 & 0 & 0 \\
0 & 0 & 1 & 0 
\end{pmatrix}
\begin{pmatrix} 
v_0 \\ v_1 \\ v_2 \\ v_3 
\end{pmatrix} ~,\\
R(e_2) X &= e_2 \wedge X = v_0 e_2 - v_1 e_1 \wedge e_2\ , \\
&= 
\begin{pmatrix} 1 & e_1 & e_2 & e_1 \wedge e_2 \end{pmatrix}
\begin{pmatrix} 
0& 0 & 0 & 0 \\
0 & 0 & 0 & 0 \\
1 & 0 & 0 & 0 \\
0 & -1 & 0 & 0 
\end{pmatrix}
\begin{pmatrix} 
v_0 \\ v_1 \\ v_2 \\ v_3 
\end{pmatrix} ~,
\end{align}
\begin{align}
R(e_1 \wedge e_2) X &=  e_1 \wedge e_2 \wedge X = v_0 e_1 \wedge e_2\ , \\
&=
\begin{pmatrix} 1 & e_1 & e_2 & e_1 \wedge e_2 \end{pmatrix}
\begin{pmatrix} 
0& 0 & 0 & 0 \\
0 & 0 & 0 & 0 \\
0 & 0 & 0 & 0 \\
1 & 0 & 0 & 0 
\end{pmatrix}
\begin{pmatrix} 
v_0 \\ v_1 \\ v_2 \\ v_3 
\end{pmatrix} 
~.
\end{align}
Here, two matrices 
\begin{align}
F_1 &= \begin{pmatrix} 
0& 0 & 0 & 0 \\
1 & 0 & 0 & 0 \\
0 & 0 & 0 & 0 \\
0 & 0 & 1 & 0 
\end{pmatrix}~,&
F_2 &= \begin{pmatrix} 
0& 0 & 0 & 0 \\
0 & 0 & 0 & 0 \\
1 & 0 & 0 & 0 \\
0 & -1 & 0 & 0 
\end{pmatrix}\ ,
\end{align}
are the representation matrices of the basis $e_1$ and $e_2$. 
The multiplication relations of $F_1$ and $F_2$ are 
\begin{align}
F_1F_2 &= 
 \begin{pmatrix} 
0& 0 & 0 & 0 \\
0 & 0 & 0 & 0 \\
0 & 0 & 0 & 0 \\
1 & 0 & 0 & 0 
\end{pmatrix}~,&
F_1F_2+F_2F_1 
 &=0~.
\end{align}
These indicate that $F_1$ and $F_2$ anticommute with each other, i.e., $F_1F_2=-F_2F_1$, 
and it can be easily confirmed that $F_1^2=F_2^2=0$. 

Moving on to the Hermitian conjugates of  $F_1$ and $F_2$, it is apparent 
that the inner product on the algebra $A$ is equivalent to the standard Hermitian inner product on $\mathbb{C}^4$. 
Therefore, the Hermitian conjugation of the linear transformation on $A$ can be obtained by taking the 
complex conjugates and transpositions of  $F_1$ and $F_2$. We obtain
\begin{align}
F_1^{\dag} &= 
\begin{pmatrix} 
0& 1 & 0 & 0 \\
0 & 0 & 0 & 0 \\
0 & 0 & 0 & 1 \\
0 & 0 & 0 & 0 
\end{pmatrix}~,&
F_2^{\dag} &= 
\begin{pmatrix} 
0& 0 & 1 & 0 \\
0 & 0 & 0 & -1 \\
0 & 0 & 0 & 0 \\
0 & 0 & 0 & 0 
\end{pmatrix}\ .
\end{align}
Analysis of the multiplicative relation between $F_1,F_2, F_1^{\dag},$ and $F_2^{\dag}$ reveals that 
the algebra is not a Grassmann algebra but is actually a Clifford algebra, with 
\begin{align}
F_1 F_1^{\dag} + F_1^{\dag} F_1 =
F_2 F_2^{\dag} + F_2^{\dag} F_2 =1\ .
\end{align}
Equivalently, 
$\{ F_1^{\dag} , F_1 \} = \{ F_2^{\dag} , F_2 \} =   1$. 
$F_1,F_2,F_1^{\dag},$ and $F_2^{\dag}$ should be treated as ``Grassmann odd quantities", 
but they do not simply anticommute with each other. 
The remaining multiplication relations are  $F_1^{\dag} F_2+F_2 F_1^{\dag}=0$ and the Hermitian conjugation. 
We obtain
\begin{align}
\{ F_i , F_j \} = \{ F_i^{\dag} , F_j^{\dag} \} = 0, ~\{ F_i , F_j^{\dag} \} = \delta_{ij} ~.
\end{align}
By taking linear combinations of $F$s, the generators $\gamma_a^{\dag}=\gamma_a,(a=1,2,3,4)$ 
of the Clifford algebra are obtained, with 
\begin{align}
\gamma_1 &= F_1+F_1^{\dag},&
 \gamma_2&= i (F_1 - F_1^{\dag} ), &
 \gamma_3 &= F_2 + F_2^{\dag},&
 \gamma_4 &= i ( F_2 - F_2^{\dag} )~,
\end{align}
and the multiplication relations of the $\gamma$'s and their Hermitian conjugates are 
\begin{align}
\{ \gamma_a , \gamma_b \} &= 2 \delta_{ab}~, &\gamma_a^{\dag} &= \gamma_a~.
\end{align}
It has therefore been shown that the extension of a Grassmann algebra with its Hermitian conjugates results in a Clifford algebra. Clearly, this result strongly suggests that we reconsider the anti-commutation relations between 
``Grassmann numbers" and their Hermitian conjugates.

\section{Discussion}

~~The construction of the inner product in the previous section is obtained 
using the Hodge dual operator.  
As shown, the algebra closed under the Hermitian conjugation is a Clifford algebra 
rather than a Grassmann algebra. 
As Clifford algebras are closely related to rotations, fermions may be related to 
rotations of certain infinite dimensional spaces, such as state spaces. 

The Grassmann algebra, $A=\mathbb{C} \oplus V \oplus ( V \wedge V)$, 
constructed from a two-dimensional 
vector space, $V$, over the complex number field, $\mathbb{C}$, is split into two parts: 
``the bosonic part", $A_0=\mathbb{C}  \oplus ( V \wedge V)$, 
and ``the fermionic part", $A_1=V$. 
For $v,w \in A_1$, $v \wedge w = - w \wedge v$. 
Let us consider $n$ ``Grassmann numbers", $( \theta_1 , \theta_2 , \cdots , \theta_n)$.
If all $\theta_i$ are elements of $A_1$ and the multiplication of 
$\theta_i$ is identified with the wedge product of the Grassmann algebra, 
all $\theta_i$ satisfy the condition that $\theta_i \theta_j + \theta_j \theta_i =0$. 
Of course, one can consider various Grassmann algebras constructed from 
 several vector spaces, and 
certain properties of $\theta_i$ depend on the dimensions of the generating vector spaces. 
For example, in the case of ${\rm dim}_{\mathbb C} V=2$, the condition, $\theta_i \theta_j \theta_k=0$, 
is satisfied for every ``configuration $(\theta_1 , \theta_2 , \cdots , \theta_n)$", whereas for ${\rm dim}_{\mathbb C} V=3$, configurations with $\theta_i \theta_j \theta_k \neq 0$ ($n>2$) exist. 
An essential question is posed as to whether ``Grassmann numbers" 
are elements of any of these possible Grassmann algebras. To answer this, we begin with the definition of a Grassmann algebra,
as we know that the dimension of the generating vector spaces is required. 

It is an assumption that the algebra, $A$, acts on the Hilbert space ${\cal H}$ as, 
without this condition, the multiplication of ``Grassmann numbers" $\theta_1, \cdots , \theta_n$
 with any states cannot be considered from the outset. 
In other words, it is conjectured that a representation $R: A \rightarrow gl({\cal H})$ is given. 
The conjecture is appropriate. 
%%%
This conjecture is equivarent to the assumption where supersymmetry is considered 
to be a symmetry of the Hilbert space. 
Let $X=\epsilon Q+ h.c.$ and $U={\rm exp}(X)=1+X+X^2/2+\cdots$. 
The action of $U, X$ and $Q^{\alpha}$ on any states $|\psi \rangle$ should be considered under the assumption;
 $U | \psi  \rangle , X | \psi  \rangle , Q^{\alpha}  | \psi  \rangle$. 
$X | \psi  \rangle = \epsilon_{\alpha} Q^{\alpha}  | \psi  \rangle$ implies 
that the multiplication of ``Grassmann numbers" $\epsilon_{\alpha}$ on the state is performed in this expression. 
 Otherwise nobody can consider the transformation of states by those supersymmetries.  
 Hence people who claim that the multiplication of ``Grassmann numbers" $\theta_1, \cdots , \theta_n$
 with any states cannot be considered from the outset are also claiming that supersymmetries are not symmetries of the 
 Hilbert space. 

Constructing spaces which have the action of a Grassmann algebra is easy. Let us show that below. 
%%%
For any vector space ${\cal H}$ over $\mathbb{C}$, by considering the extension of the coefficients 
\begin{align}
{\cal H}_{\theta} &= A \otimes_{\mathbb{C}} {\cal H}~~,
\end{align}
we obtain a vector space ${\cal H}_{\theta}$, which has the action of $A$.
The set, ${\cal H}_{\theta}$, is actually a vector space over $\mathbb{C}$, 
and it is easily shown that every finite-dimensional vector space over $\mathbb{C}$ has 
a positive definite Hermitian inner product. 
If there is a positive definite inner product on $V$ which generates $A$, 
a positive definite inner product on $A$ can be defined using the Hodge dual operator. 
Thus, ${\cal H}_{\theta}$ becomes a vector space with a positive definite inner product.

\section{Conclusion}
This paper concludes that the property stating that a ``Grassmann number", 
$\epsilon$, commutes with ``bosonic quantities" and anticommutes with ``fermionic quantities" 
is not appropriate to define the idea of ``Grassmann numbers" and, as a result, 
doubt is cast on all calculations involving Grassmann numbers.  
The property is different from that of Grassmann algebra and, in particular, the foundation of supersymmetry must be reconsidered.

\section*{Acknowledgements}
The author appreciates Doctor Taro Kashiwa and Doctor Ryuichiro Kitano's assistance in reading the draft paper and in supplying advice. 
The author thanks Doctor Taichiro Kugo for recommending publication of this paper.  
The author would like to express his gratitude to Alexandre Kabbach for proofreading the draft.  

I will dedicate this paper to my sister, Yuki Kihara, in heaven.

\end{document}